\documentclass[aps,twocolumn,preprintnumbers,amsmath,amssymb,superscriptaddress,floatfix,nofootinbib]{revtex4}

\usepackage{graphicx}
\usepackage{epsfig}
\usepackage{epstopdf}
\usepackage{hyperref}
\usepackage{amsmath}
\usepackage{amsfonts}
\usepackage{amssymb}
\usepackage{color}
\usepackage{subfigure}
\usepackage{CJKutf8}
\begin{document}
\begin{CJK}{UTF8}{gbsn}
\title{Explore the properties of $\Lambda(1670)$ in the Cabibbo-favored process $\Lambda^+_c \to p K^- \pi^+$}

\author{Sheng-Chao Zhang}
\affiliation{School of Physics and Electronics, Henan University, Kaifeng 475004, China}
\author{Man-Yu Duan}
\affiliation{School of Physics, Southeast University, Nanjing 210094, China}
\affiliation{School of Physics, Zhengzhou University, Zhengzhou 450001, China}
\affiliation{Departamento de Física Teórica and IFIC, Centro Mixto Universidad de Valencia-CSIC Institutos de Investigación de Paterna, 46071 Valencia, Spain}

\author{Wen-Tao Lyu}
\affiliation{School of Physics, Zhengzhou University, Zhengzhou 450001, China}
\author{Guan-Ying Wang}\email{wangguanying@henu.edu.cn}
\affiliation{School of Physics and Electronics, Henan University, Kaifeng 475004, China}

\author{Jing-Yu Zhu}\email{zhujingyu@zzu.edu.cn}
\affiliation{School of Physics, Zhengzhou University, Zhengzhou 450001, China}

\author{En Wang}
\affiliation{School of Physics, Zhengzhou University, Zhengzhou 450001, China}

\date{\today}

\begin{abstract}

Recently, the Belle and LHCb Collaborations have measured the $\Lambda^+_c \to p K^- \pi^+$ decay and reported the $K^- p$ invariant mass distribution, which shows a clear cusp structure around  the $\eta \Lambda$ threshold. In this work, we have analyzed this process by considering the triangle mechanism and the $S$-wave pseudoscalar meson-octet baryon interactions within the chiral unitary approach, which dynamically generates the $\Lambda(1670)$. 
Our results are in good agreement with the Belle measurements, which implies that the cusp structure around $\eta\Lambda$ threshold could be associated with the $\Lambda(1670)$ with the molecular nature. 
\end{abstract}

\maketitle

\section{Introduction}
Since the discovery of the charmed baryon $\Lambda_c^+$~\cite{Knapp:1976qw}, there has been many studies of the hadronic weak decays of charmed baryons, which provide an important platform to study the strong and weak interactions~\cite{Miyahara:2015cja,Jia:2024pyb,Geng:2018upx,Geng:2024sgq}. Yet, both the experimental and theoretical progresses in the study of hadronic decays of charmed baryons were very slow for a long time. This situation was reversed since 2014 as there were several major breakthroughs in charmed-baryon experiments~\cite{Cheng:2021qpd,Zhong:2024zme}. 
In the last decade, there are a lot of multi-body hadronic decay modes of the charmed baryons reported by the BESIII, Belle, and LHCb Collaborations~\cite{ParticleDataGroup:2024cfk,BESIII:2018qyg,BESIII:2023sdr,Belle:2022pwd,LHCb:2017xtf,LHCb:2019ldj},  which involve  complicated final state interactions, and have motivated many studies of the light hadrons in those processes, such as the scalar mesons $a_0(980)/f_0(980)$ in $\Lambda^+_c \to p K^+ K^-, p \pi^+ \pi^-$~\cite{Wang:2020pem}, $\Lambda^+_c \to p \eta \pi^+$~\cite{Feng:2020jvp}, and light baryons in $\Lambda^+_c \to {\pi}^0 {\phi} p$~\cite{Xie:2017mbe}, $\Lambda^+_c \to \bar{K}^0 \eta p$~\cite{Xie:2017erh,Li:2024rqb}, $\Lambda^+_c \to \Lambda \eta \pi^+$~\cite{Wang:2022nac,Xie:2017xwx,Xie:2016evi,Lyu:2024qgc}, $\Lambda^+_c \to p K^- \pi^+$~\cite{Liu:2019dqc}.

 In 2023, the LHCb Collaboration has performed an amplitude analysis of the process $\Lambda^+_c \to p K^- \pi^+$ decay in connection with the semi-leptonic beauty hadron decays~\cite{LHCb:2022sck}, where a cusp structure around $\eta\Lambda$ threshold appears in the measured  $K^- p$ invariant mass distribution. Later, the Belle Collaboration has also measured the process $\Lambda^+_c \to p K^- \pi^+$, and found a narrow peaking structure near the $\Lambda \eta$ threshold in the $ K^-p$ invariant-mass distribution~\cite{Belle:2022cbs}. The Belle has explained this structure within the two approaches, the Breit-Wigner function of a new resonance, and  a Flatt\'e function, and suggested that the cusp structure around the $\eta\Lambda$ could be associated with $\Lambda(1670)$~\cite{Belle:2022cbs}. Indeed,  the process $\Lambda^+_c \to p K^- \pi^+$ has been studied for a long time, and  its complex resonant structure has attracted widespread attention from theorists and experimentalists~\cite{Filaseta:1987rx,BESIII:2015bjk,Liu:2019dqc,Miyahara:2015cja,Duan:2024okk}. 
    
  In Ref.~\cite{Liu:2019dqc}, the authors assumed that  the process $\Lambda^+_c \to p K^- \pi^+$ could happen via the two triangle loops composed of the $a_0(980)^+ \Lambda$ and $\Sigma^*(1660) \eta$ intermediate states with a fusion of the $\eta\Lambda$ into the $\Lambda(1670)$, followed by the $\Lambda(1670)$ decaying into the final states $K^-p$. While these loops do not develop a triangle singularity (TS) in the physical region, the TS in the complex plane of the second Riemann sheet is close to the $\eta\Lambda$ threshold and can enhance the threshold cusp~\cite{Guo:2019twa}, providing a natural explanation of the narrow cusp structure observed in  $\Lambda^+_c \to p K^- \pi^+$~\cite{Liu:2019dqc}.  In addition, Ref.~\cite{Ahn:2019rdr} has investigated $S = -1$ hyperon production from the process $\Lambda^+_c \to p K^- \pi^+$ within the effective Lagrangian approach, and a sharp peak-like structure near 1665~MeV is predicted, mostly resulting from  the interference effects between the resonance $\Lambda(1670)$  and the $\eta$-$\Lambda$ loop channels.

As we know, the ground octet and decuplet baryons could be well described within the naive quark model, however the low-lying excited baryons with the quantum numbers of spin-parity $J^P=1/2^-$ have some exotic properties, which are difficult to be explained within the naive quark model. For instance~\cite{Wang:2024jyk}, the $\Lambda(1405)$ mass is difficult to understand in the naive quark model as a $P$-wave excited baryon~\cite{Isgur:1978xj}, and this state could be generated by the $\Bar{K}N$ coupled-channel interaction within the unitary chiral approach~\cite{Kaiser:1995eg,Oset:1997it,Oller:2000fj}. 
Concerning the low-lying excited baryon $\Lambda(1670)$, it exhibits itself as an enhancement structure in the $\eta\Lambda$ invariant mass distribution of the processes $K^-p\to \eta\Lambda$~\cite{CrystalBall:2001uhc} and $\Lambda_c^+\to \eta\Lambda\pi^+$~\cite{BESIII:2018qyg,Belle:2020xku}, and also a dip structure around $\eta\Lambda$ threshold in the $\bar{K}^0n$ invariant mass distribution of the process $K^-p\to \bar{K}^0n$~\cite{Gopal:1976gs}. Indeed, the dip structure could be produced by the interference between the resonance and the other contributions~\cite{Feng:2020jvp,Guo:2019twa,Ding:2023eps,Lu:2016roh,Chen:2015sxa}, and one nice example is that the $f_0(980)$ appears as a dip for the process $J/\psi\to \omega \pi^+\pi^-$~\cite{BES:2004mws}, but as a sharp peak in the $\pi^+\pi^-$ invariant mass distribution for the process $J/\psi\to \phi\pi^+\pi^-$~\cite{BES:2004twe}.     

The authors of Ref.~\cite{Zhong:2008km} has analyzed theoretically  the differential cross section and total cross section of the process $K^- p \to \pi^0 \Sigma^0$ within the chiral quark model and suggested that the $\Lambda(1670)$ is dominated by the configuration of $[70,^28,1/2]$.
The mass of $\Lambda(1670)$ is close to the $\eta\Lambda$ threshold, thus, suggesting  that the hadron-hadron interactions play an important role in the $\Lambda(1670)$ dynamics structure~\cite{Dong:2021juy,Guo:2017jvc}. In  Refs.~\cite{Oset:2001cn,Garcia-Recio:2002yxy,Oller:2006jw,Kamano:2015hxa}, the $\Lambda(1670)$ could be explained as a dynamically generated resonance of the $S$-wave meson-baryon interactions within the chiral unitary approach, and this explanation is supported by the studies of Refs.~\cite{Wang:2022nac,Lyu:2024qgc,Kamano:2015hxa}.

In Ref.~\cite{Menadue:2011pd}, the authors have performed the first investigation of the low-lying $J^P=1/2^-$ spectrum of the $\Lambda$ baryon at near-physical pion masses, and identified a low-lying $\Lambda(1405)$ using conventional three-quark operators. Meanwhile, two $\Lambda$ states are found in the $\Lambda(1670)$ mass region, which implies that a single particle state makes an important role to the structure of the $\Lambda(1670)$.

As pointed out in Refs.~\cite{Dong:2020hxe,Bugg:2008wu,Guo:2014iya}, a pronounced threshold cusp requires the existence of the nearby pole, and the cusp structure observed in the $pK^-$ invariant mass distribution of the process $\Lambda^+_c \to p K^- \pi^+$ by LHCb and 
Belle should be helpful to understand the properties of the $\Lambda(1670)$ and the meson-baryon interactions. 
Thus,  in this work  we will investigate the process $\Lambda^+_c \to  p K^- \pi^+ $ by considering  the triangle mechanism and the $S$-wave meson-baryon interaction within the chiral unitary approach， which could generate the resonance $\Lambda(1670)$.

    This article is organized as follows. In Sec.~\ref{Sec:Formalism}, we present the theoretical formalism of the $\Lambda^+_c \to  p K^- \pi^+ $. Numerical results and discussions are presented in Sec.~\ref{Sec:Results}, followed by a short summary in the last section.

\section{Formalism} \label{Sec:Formalism}

In this section, we will introduce the  formalism of the triangle mechanism for the process  $\Lambda^+_c \to  p K^- \pi^+ $, and give the formalism of the invariant mass distributions for this process.

In Ref.~\cite{Liu:2019dqc}, the authors have considered two triangle loops composed of the $a_0(980)^+ \Lambda$ and $\Sigma^*(1660) \eta$ intermediate states for the process $\Lambda^+_c \to p K^- \pi^+$. However, there is no significant structure of $\Sigma^*(1660)$ in the Dalitz plot of the process $\Lambda^+_c \to \eta \Lambda \pi^+$ measured by BESIII and Belle~\cite{BESIII:2018qyg,Belle:2020xku}. Meanwhile, the process $\Lambda_c^+\to \eta \Sigma^*(1660)$ has not been observed so far. On the other hand, the $ud$ diquark in the $\Lambda_c^+$ is the most attractive ``good'' diquark with isospin $I=0$, the process $\Lambda_c^+\to \eta \Sigma^*(1660)$ will change the isospin of $ud$ diquark and  destruct the strong diquark correlation~\cite{Jaffe:2004ph,Miyahara:2015cja}. Thus, in this work, we will neglect the triangle loop with the intermediate $\Sigma^*(1660) \eta$ state, and only consider the one of $a_0(980)^+ \Lambda$.

\begin{figure}[htbp]
\begin{center}
\includegraphics[width=0.45\textwidth]{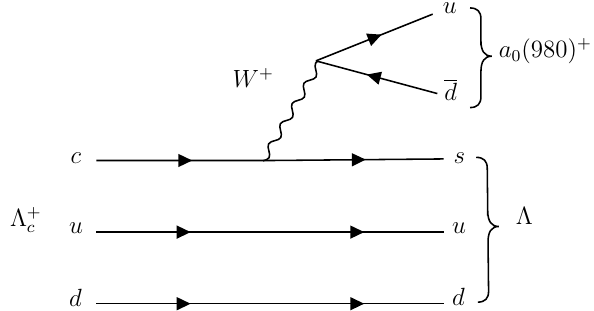}
\caption{Feynman diagram at the quark level for the weakly
process $\Lambda_{c}^{+}\rightarrow W^{+}+s+ud\rightarrow u\bar{d}+s+ud$.}
\label{Fig:wei}
\end{center}
\end{figure}

The process $\Lambda_c^+\to a_0(980)^+ \Lambda$ could happen via the following steps. First, the $c$ quark of the initial $\Lambda^+_c$ weakly decays into an $s$ quark plus a $W^+$ boson, then the $W^+$ boson goes into a $u \bar{d}$ pair, as shown in Fig.~\ref{Fig:wei}.
Second, the $u \bar{d}$ pair from the $W^+$ decay
will hadronize into $a_0(980)^+$, while the $s$ quark from the $c$ decay, and the $ud$ pair of the initial $\Lambda^+_c$, combine to give  the state $\Lambda$~\cite{Wang:2022nac,Lyu:2024qgc}.

\begin{figure}[htbp]
\begin{center}
\includegraphics[width=0.45\textwidth]{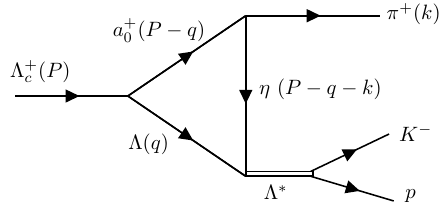}
\caption{Triangle mechanism for the process $\Lambda^+_c \to  p K^- \pi^+ $. Kinematic conventions for the intermediate states
are $\Lambda_c^+(P)$, $\Lambda(q)$, $a_0(980)(P-q)$, $\pi^+(k)$ and $\eta(P-q-k)$.} 
\label{Fig:TS-hardon}
\end{center}
\end{figure}

For the triangle loops composed of the $a_0(980)^+$ as depicted in Fig.~\ref{Fig:TS-hardon}, the $a_0(980)^+$ decays into  an internal state $\eta$ and an external state $\pi^+$. The triangle singularities appear in a particular situation, where all the intermediate states are placed on shell. In the rest frame of $\Lambda_c^+$, the momenta of the $\Lambda$ and the $\eta$ are parallel, and they could undergo the rescattering to the final states $K^-p$. The classical situation requires that the $\eta$ moves along the same direction and faster that the state $\Lambda$. This is the essence of the Coleman-Norton theorem when applied to a triangle diagram~\cite{Guo:2019twa,Coleman:1965xm,Bayar:2016ftu,Liang:2019jtr,Wang:2016dtb}.

Applying Eq.~(18) of Ref.~\cite{Bayar:2016ftu}, one can see that the TS appears if the mass of the $\Lambda_c^+$ is 2128~MeV, 160~MeV lower than the $\Lambda_c^+$ nominal mass. Conversely, if we take the physical mass of the $\Lambda_c^+$, the TS appears for an $a_0$ mass of 1105~MeV, again 125~MeV away from the nominal $a_0(980)$ mass. This situation is far away to have large effects from a TS, in spite of which some enhancement is still found in Ref.~\cite{Liu:2019dqc}. 

Firstly, we consider the decay of $\Lambda_c^+ \to \Lambda a_0(980)$ via $S$-wave, and the amplitude could be given as,
\begin{equation}
	t_{\Lambda_c^+\to \Lambda a_0(980)^+}=g_{\Lambda_c^+a_0\Lambda},
\end{equation}
where the constant $g_{\Lambda_c^+a_0\Lambda}$ corresponds
to the effective coupling between $\Lambda_c^+$ and $a_0(980)^+\Lambda$. And the amplitude for the vertex of  $a_0(980)^+\to \eta\pi^+$ can be written as,
\begin{equation}
	t_{a_0(980)^+\to \eta\pi^+}=g_{a_0\eta\pi},
\end{equation}
with the effective coupling $g_{a_0\eta\pi}$ between $a_0(980)^+$ and $\eta\pi^+$.

Then, we will calculate the transition amplitude for the $S$-wave $\Lambda\eta\to K^-p$ interaction, which can be obtained by solving the Bethe-Salpeter (BS) equation,
\begin{equation}\label{BS}
	T=[1-VG]^{-1}V,
\end{equation}
with the potential $V_{ij}$ between the coupled channels ~\cite{Oset:2001cn},
\begin{equation}\label{V}
	\begin{aligned}
		V_{ij} &=-C_{ij}\frac{1}{4f^2}(2\sqrt{s}-M_i-M_j)\\
		&\times\left(\dfrac{M_i+E_i}{2M_i}\right)^{1/2}\left(\dfrac{M_j+E_j}{2M_j}\right)^{1/2},
	\end{aligned}
\end{equation}
where $E_i$ and $M_i$ are the energy and mass of the baryon in the $i$-th channel, and the coefficients $C_{ij}$ reflect the $SU(3)$ flavor symmetry, and are given in Table~I of Ref.~\cite{Oset:1997it}. Here we have considered the coupled channels of $\bar{K}N$, $\pi\Sigma$, $\eta\Lambda$, $K\Xi$. $\sqrt{s}$ is the invariant mass of the $K^-p$ system. The coupling $f$ is the pseudoscalar decay constant and we use,
\begin{equation}
	f=1.15f_{\pi},~~f_{\pi}=93~\mathrm{MeV}.
\end{equation}
The meson-baryon loop function $G_{l}$ can be written by~\cite{Oset:2001cn},
\begin{equation}\label{G-DR}
	\begin{aligned}
		G_{l}  =&i \int \frac{d^4 q}{(2 \pi)^4}\frac{M_l}{E_l(\vec{q}\,)} \frac{1}{\sqrt{s}-q^0-E_l(\vec{q}\,)+i \epsilon} \frac{1}{q^2-m_l^2+i \epsilon} \\
		 =&\frac{2M_l}{16 \pi^2}\left\{a_l(\mu)+\ln \frac{M_l^2}{\mu^2}+\frac{s+m_l^2-M_l^2}{2 s} \ln \frac{m_l^2}{M_l^2}\right. \\
		& +\frac{|\vec{q}\,|}{\sqrt{s}}\left[\ln \left(s-\left(M_l^2-m_l^2\right)+2 |\vec{q}\,| \sqrt{s}\right)\right. \\
		& +\ln \left(s+\left(M_l^2-m_l^2\right)+2 |\vec{q}\,| \sqrt{s}\right) \\
		& -\ln \left(-s+\left(M_l^2-m_l^2\right)+2 |\vec{q}\,| \sqrt{s}\right) \\
		& \left.\left.-\ln \left(-s-\left(M_l^2-m_l^2\right)+2 |\vec{q}\,| \sqrt{s}\right)\right]\right\},
	\end{aligned}
\end{equation}
where $m_l$ and $M_l$ are the masses of meson and baryon of the $l$-th channel. Here, we take $\mu=630$~MeV\footnote{The changes in the $\mu$ can be accommodated in terms of changes in the subtraction constant $a_l$~\cite{Oset:2001cn}.}. Since the pole position of $\Lambda(1670)$ is quite sensitive to the value of $a_{K\Xi}$ and only moderately sensitive to the subtraction constant $a_{\bar{K}N}$, $a_{\pi\Sigma}$, $a_{\eta\Lambda}$, we take $a_{K\Xi}$ to be a free parameter, and adopt the values $a_{\bar{K}N}=-1.84$, $a_{\pi\Sigma}=-2.00$, and $a_{\eta\Lambda}=-2.25$ from Refs.~\cite{Oset:2001cn,Wang:2022nac,Lyu:2024qgc}.  

Now, having the above amplitudes,  we could write the total invariant decay amplitude of Fig.~\ref{Fig:TS-hardon},
\begin{eqnarray}\label{Eq:t-TS}
	\mathcal{T}^{\rm TS}=\mathcal{Q}t^{TS} \times t_{\Lambda\eta\to K^-p}, 
\end{eqnarray}
where $\mathcal{Q}=g_{\Lambda_c^+a_0\Lambda}g_{a_0\eta\pi}$ is a constant. $t^{TS}$ is the triangle amplitude of Fig.~\ref{Fig:TS-hardon}, and  is given by, 
\begin{align}\label{Eq:TS}
	t^{TS} =&i\int\frac{d^4q}{(2\pi)^4} \dfrac{2m_\Lambda}{q^2-m_\Lambda^2+i\varepsilon} \dfrac{1}{(P-q)^2-m_{a_0}+im_{a_0}\Gamma_{a_0}}  \nonumber\\
	&\times \dfrac{1}{(P-q-k)^2-m_{\eta}^2+i\varepsilon} ,
\end{align}
by performing the $q^0$ integration analytically we finally get an easy expression,
\begin{align}
	t^{TS} &=\int\frac{d^3q}{(2\pi)^3} \dfrac{2m_\Lambda}{8\omega_\Lambda\omega_{a_0}\omega_{\eta}} \dfrac{1}{k^0-\omega_{a_0}-\omega_{\Lambda}+i\frac{\Gamma_{a_0}}{2}}  \nonumber\\
	&\times \dfrac{1}{P^0+\omega_\Lambda+\omega_{\eta}-k^0}   \nonumber\\
	&\times \dfrac{2P^0\omega_\Lambda+2k^0\omega_{\eta}-2(\omega_\Lambda+\omega_{\eta})(\omega_\Lambda+\omega_{\eta}+\omega_{a_0})}{P^0-\omega_{a_0}-\omega_{\Lambda}+i\frac{\Gamma_{a_0}}{2}}  \nonumber\\
	&\times \dfrac{1}{P^0-\omega_\Lambda-\omega_{\eta}-k^0+i\varepsilon}, \label{eq:TSamp}
\end{align}
with $P^0=M_{\Lambda_c^+}$, $\omega_\Lambda=\sqrt{|\vec{q}\,|^2+m_{\Lambda}^2}$, $\omega_{\eta}=\sqrt{|\vec{q}+\vec{k}|^2+m_{\eta}^2}$, and $\omega_{a_0}=\sqrt{|\vec{q}\,|^2+m_{a_0}^2}$. Here $m_{a_0}=980$~MeV, and $\Gamma_{a_0}=75$~MeV~\cite{ParticleDataGroup:2024cfk,Liu:2019dqc}\footnote{As we know, the width of the scalar $a_0(980)$ has not been well established due to its exotic structure, thus we take the centre value of the estimated range $50\sim 100$~MeV of Particle Data Group (PDG)~\cite{ParticleDataGroup:2024cfk}. Indeed, it has been shown in Ref.~\cite{Liu:2019dqc} that the width of $a_0(980)$ has less influence on the contribution from the triangle mechanism.}.  $k^0$ and  $\vec{k}$ are the energy and three-momentum of the $\pi^+$ emitted from $a_0(980)$ in the rest frame of the $\Lambda_c^+$, which are given by
\begin{equation}
	k^0=\dfrac{M_{\Lambda_c^+}^2+m_{\pi^+}^2-m_{K^-p}^2}{2M_{\Lambda_c^+}},
\end{equation}
\begin{equation}
	|\vec{k}|=\sqrt{|k^0|^2-m_{\pi^+}^2}.
\end{equation}

In this work, to regularize the integral of Eq.~(\ref{eq:TSamp}), we use the cutoff $q_{\rm max}=1000$~MeV, and will show that our results are insensitive to its value.

With the amplitude of Eq.~(\ref{Eq:t-TS}), we can write the $K^-p$ invariant mass distribution for the process  $\Lambda_c^+\to p K^-\pi^+$,
\begin{equation}
\frac{d\Gamma}{dM_{K^-p}}=\frac{1}{(2\pi)^3}\frac{2M_{p}2M_{\Lambda_c}p_{\pi^+} \tilde{p}_{K^{-}}}{4M^2_{\Lambda_c^+}}|{\mathcal{T}^{\rm TS}}|^2,
\label{eq:dw_threebody}
\end{equation}
where $p_{\pi^+}$ is the $\pi^+$ momentum in the $\Lambda_{c}^+$ rest frame,
\begin{equation}
p_{\pi^+}=\frac{\lambda^{1/2}(M^2_{\Lambda_c^+}, M^2_{\pi^+}, M^2_{K^-p}
)}{2M_{\Lambda_c^+}},
\end{equation}
$\tilde{p}_{K^-}$ is the $K^-$ momentum in the $K^- p$ rest frame.
\begin{eqnarray}
\tilde{p}_{K^-} =
\frac{\lambda^{1/2}(M^2_{K^-p}, m^2_{K^-}, m^2_{p})}{2M_{K^-p}},
\end{eqnarray}
with K{\"a}llen function $\lambda(x,y,z)=x^2+y^2+z^2-2xy-2yz-2zx$. All the masses of the mesons and baryons involved in our calculations are taken from {PDG}~\cite{ParticleDataGroup:2024cfk}.

\section{Numerical results and discussion} \label{Sec:Results}

\begin{figure}[htbp]
\begin{center}
\includegraphics[width=0.45\textwidth]{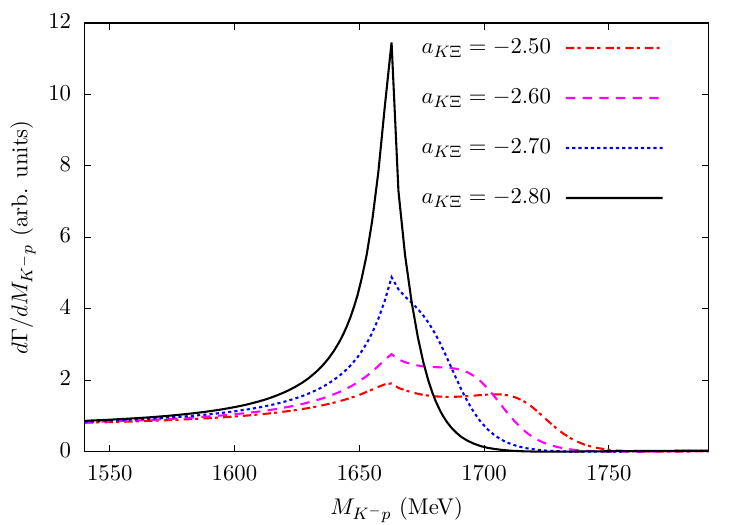}
\caption{(Color online) Invariant $ K^-p $ mass distribution
for the $\Lambda^+_c \to p K^- \pi^+$ decay considering the triangle mechanism.}
\label{fig:pk-TS}
\end{center}
\end{figure}

\begin{figure}[htbp]
\begin{center}
\includegraphics[width=0.45\textwidth]{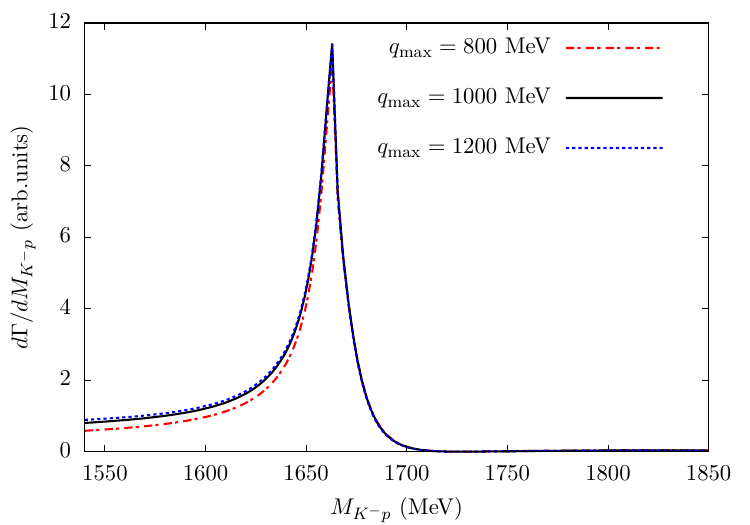}
\caption{(Color online) Invariant $K^- p$ mass distribution
for the $\Lambda^+_c \to p K^- \pi^+$ decay considering the triangle mechanism with $q_{\rm max}=800, 1000, 1200$~MeV.}
\label{fig:pk_qmax}
\end{center}
\end{figure}

\begin{table}[htpb]
	\begin{center}	
		\caption{\label{tab:1}The fitted parameter values in this work.}
		\begin{tabular}{cc}
			\hline\hline
		   parameters      & values \\
			\hline
			$\mathcal{Q}'$       &$(6.092\pm0.068)\times10^{4}$ \\
			$a_{K\Xi}$  &  $-2.776\pm0.003$ \\
			$a_0$       &   $-1969.30\pm0.23 $\\
            $a_1$       &$3.860\pm0.001$ \\
            $a_2$       &$(-2.455\pm0.001)\times10^{-3}$ \\
			$a_3$       & $(5.118\pm0.001)\times10^{-7}$ \\
            \hline
            $\chi^2$   &$301.76$   \\
            $\chi^2/d.o.f.$  &$1.25$   \\
			\hline\hline
		\end{tabular}
	\end{center}
\end{table}

\begin{figure}[htbp]
\begin{center}
\includegraphics[width=0.45\textwidth]{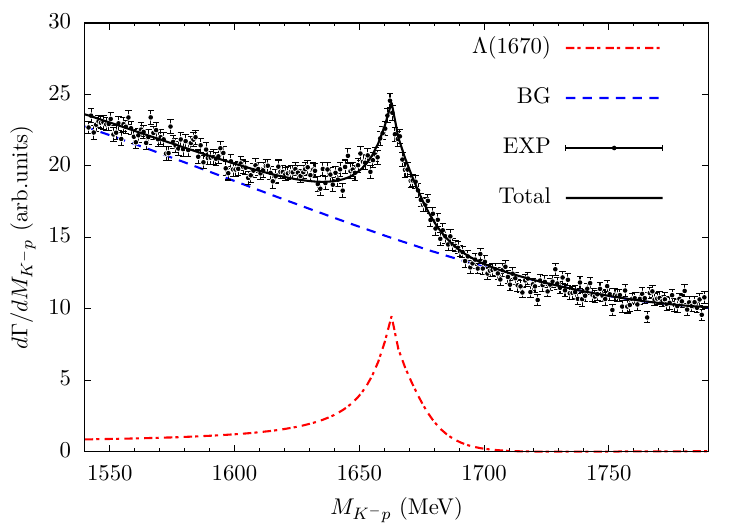}
\caption{(Color online) Invariant $K^- p$ mass distribution
for the $\Lambda^+_c \to p K^- \pi^+$ decay considering the triangle mechanism with fitted parameters.}
\label{fig:pk-TS-fit}
\end{center}
\end{figure}

In our model, we have two free parameters, the subtract constant $a_{K\Xi}$, related to the pole position of the $\Lambda(1670)$, and the coupling constant $\mathcal{Q}=g_{\Lambda_c^+a_0\Lambda}g_{a_0\eta\pi}$ of Eq.~(\ref{Eq:t-TS}), which is a normalization factor. Up to an arbitrary normalization, we show the $K^- p$ invariant mass distribution of the process $\Lambda^+_c \to p K^- \pi^+$ with different values of the subtraction constants $a_{K\Xi}=-2.50$, $-2.60$, $-2.70$, and $-2.80$ in Fig.~\ref{fig:pk-TS}. One can find that, the pole position of the $\Lambda(1670)$ is quite sensitive to the value of the  subtraction constants $a_{K\Xi}$. For  larger values of $a_{K\Xi}=-2.50, -2.60$,  there is a cusp structure around the $\eta\Lambda$ threshold followed by a bump structure, and for  smaller values of $a_{K\Xi}=-2.70, -2.80$, there is a clear cusp structure around the $\eta\Lambda$ threshold.  Thus, the cusp structure observed in the $K^-p$ invariant mass distribution of the process  $\Lambda^+_c \to p K^- \pi^+$ by Belle should be associated with the $\Lambda(1670)$~\cite{Belle:2022cbs}, and the lineshape should be helpful to constrain the value of the subtraction constants $a_{K\Xi}$.

Meanwhile, in Fig.~\ref{fig:pk_qmax} we also show the results with $a_{K\Xi}=-2.80$ and different values of the cut-off of Eq.~(\ref{eq:TSamp}), i.e. $q_{\rm max}=800, 1000, 1200$~MeV, and one can find that the line shape and the cusp position have no sizeable changes. Thus, in the following we set $q_{\rm max}=1000$~MeV.

In order to fit to the $K^-p$ invariant mass distribution of the efficiency-corrected events measured by Belle~\cite{Belle:2022cbs},
we have added incoherently background to the TS contribution, and adopted a polynomial $f(M_{K^-p})=a_0+a_1M_{K^-p}+a_2M_{K^-p}^2+a_3M_{K^-p}^3$ to account for the background, where the $a_0$, $a_1$, $a_2$, and $a_3$ are the free parameters. In addition, the coupling constant $\mathcal{Q}$ should be replaced by a normalization factor $\mathcal{Q}'$.  In Fig.~5 of Belle measurements~\cite{Belle:2022cbs}, there are 248 data points in total. We have tabulated the fitted parameters in Table~\ref{tab:1}, and obtained the fitted $\chi^2/N_{d.o.f.}=1.25$. Here the fitted subtract constant $a_{K\Xi}=-2.776\pm 0.003$, which gives rise to a cusp structure of $\Lambda(1670)$ in the $K^-p$ invariant mass distribution.
With the fitted parameters, we have calculated the invariant mass distributions of $K^-p$, as shown in Fig.~\ref{fig:pk-TS-fit}. The red-dashed-dotted curve shows the contribution from the resonance $\Lambda(1670)$ and the triangle mechanism,  the blue-dashed curve shows the background contribution of the polynomial $f(M_{K^-p})$,  and the black-solid shows the total results. The experimental data labeled by `EXP' are take from the Fig.~5 of Ref.~\cite{Belle:2022cbs}.
One can find that, our results are in good agreement with the Belle measurements. The cusp around $\eta\Lambda$ threshold in $K^-p$ mass invariant distribution observed by Belle could be associated with the resonance $\Lambda(1670)$, and triangle mechanism could enhance the threshold cusp, as pointed in Ref.~\cite{Liu:2019dqc}.

\begin{figure}[htbp]
\begin{center}
\includegraphics[width=0.45\textwidth]{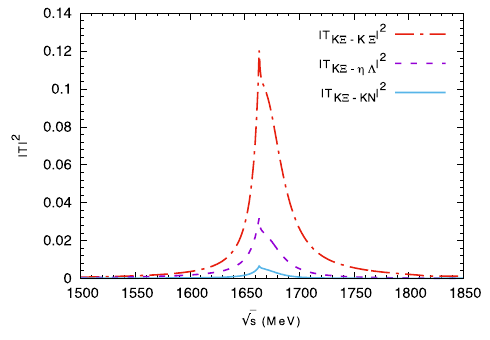}
\caption{(Color online) Modulus squared of the transition amplitudes $\left|T_{K\Xi-K\Xi}\right|^{2}$, $\left|T_{K\Xi-\eta\Lambda}\right|^{2}$, $\left|T_{K\Xi-KN}\right|^{2}$, respectively.} 
\label{Fig:T2}
\end{center}
\end{figure}

With the value of $a_{K\Xi}=-2.776$, we have presented the modules squared of the transition amplitudes $\left|T_{K\Xi-K\Xi}\right|^{2}$, $\left|T_{K\Xi-\eta\Lambda}\right|^{2}$, $\left|T_{K\Xi-KN}\right|^{2}$ in Fig.~\ref{Fig:T2}, where one can find that the position of the pole is $(1669.2,20.4i)$~MeV, in consistent with the results of Ref.~\cite{Kamano:2015hxa}.

\begin{figure*}[htbp]
    \hspace{-9cm}
    \begin{minipage}{8cm}
    \includegraphics[width=8cm]{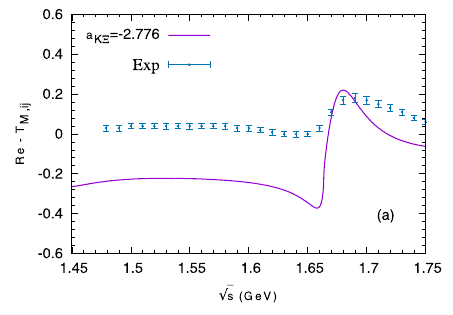}
    \end{minipage}
    \begin{minipage}{-6cm}
    \hspace{-5cm}
    \includegraphics[width=8cm]{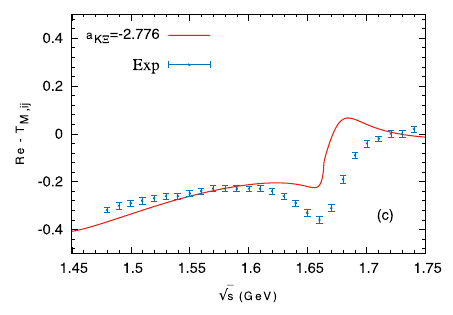}
    \end{minipage}

    \hspace{-9cm}
    \begin{minipage}{8cm}
    \includegraphics[width=8cm]{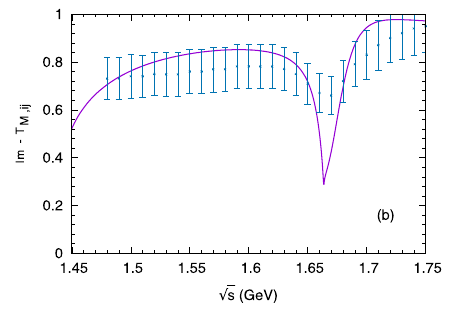}
    \end{minipage}
    \begin{minipage}{-1.0cm}
    \hspace{-5cm}
    \includegraphics[width=8cm]{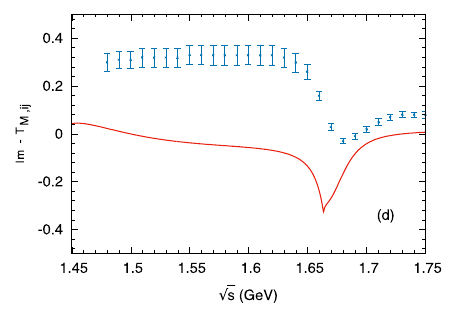}
    \end{minipage}
    \caption{(Color online)The figures (a) and (b) are the real part  and imaginary part  of the  $\bar{K}N \to \bar{K}N$ amplitude, while the figures (c) and (d) are the real part and imaginary part of $\bar{K}N \to \pi \Sigma$ amplitude. Here $a_{K\Xi}=-2.776$ and the experimental data are taken from Ref.~\cite{Gopal:1976gs}.  }
\label{fig:oset}
\end{figure*}

Furthermore, with $a_{K\Xi}=-2.776$ we have presented the real and imaginary parts of the transition amplitudes of $\bar{K}N\to \bar{K}N$ and $\bar{K}N\to \pi\Sigma$ in Fig.~\ref{fig:oset}, in order to compare with the experimental results~\cite{Gopal:1976gs}. It should be pointed out that, the normalization of the amplitudes shown in Fig.~\ref{fig:oset} is different with the one of Eq.~(\ref{BS}), and the plotted amplitudes $T'_{ij}$ can be related to the ones of Eq.~(\ref{BS}) by following relationship~\cite{Oset:2001cn,Gopal:1976gs},
\begin{equation}
    T'_{ij}=-T_{ij}\frac{\sqrt{M_iM_jp_ip_j}}{4\pi\sqrt s},
\end{equation}
where $M_{i,j}$ and $p_{i,j}$ are the masses and three-momentum modulus of baryons in the $i(j)$-th coupled channel, respectively, and the $\sqrt{s}$ is the invariant mass of the meson-baryon system. 

One can find that, our results are in agreement with the imaginary part of the $\bar{K}N\to \bar{K}N$ transition amplitude within uncertainties, while our prediction for the real part below the $\eta\Lambda$ threshold differs from the data~\cite{Gopal:1976gs}, which implies that a large constant background contribution should be taken into account~\cite{Oset:2001cn}. 
Our results of the $\bar{K}N\to \pi\Sigma$ transition amplitude show the resonant feature with the same pattern as the experiment~\cite{Gopal:1976gs}. The disagreement of the imaginary part with the data also implies an apparent background contributions missed in our model~\cite{Oset:2001cn},  such as the $t/u$-channel exchange terms.

\section{Summary}
Motivated by the cusp structure around $\eta\Lambda$ threshold in the $K^-p$ invariant mass distribution of the process $\Lambda^+_c \to p K^- \pi^+$ observed by the LHCb and Belle Collaborations, we have investigated this process by taking into account triangle mechanism and the $S$-wave meson-baryon interactions within the chiral unitary approach, which could dynamically generate the resonance $\Lambda(1670)$.  The process  $\Lambda^+_c \to p K^- \pi^+$ could happen via the triangle loop of $a_0(980)^+\Lambda$ intermediate, followed by the $a_0(980)^+$ decaying into $\eta$ and $\pi^+$, then the fusion of $\eta$ and $\Lambda$ gives rise to the $\Lambda(1670)$, following by the $\Lambda(1670)$ decaying into $K^-p$. 

We have shown that the pole position of the $\Lambda(1670)$ is quite sensitive to the subtract constant $a_{K\Xi}$ that regularizes the meson-baryon loop functions. By fitting  to the $K^-p$ invariant mass distribution of the efficiency-corrected events measured by Belle, we have obtained the $\chi^2/N_{d.o.f.}=1.25$, and the fitted subtract constant $a_{K\Xi}=-2.776\pm 0.003$, which corresponds to the pole of $(1669.0,20.3i)$~MeV for the $\Lambda(1670)$. Our results are in good agreement with the $K^-p$ invariant mass distribution of the Belle data, which supports that the observed cusp structure around $\eta\Lambda$ threshold could be associated with the $\Lambda(1670)$.

In order to extract the accurate properties of the $\Lambda(1670)$, one could search for this state in the processes with more kinematic phase space for the $\Lambda(1670)$. It should be stressed that, the process $\Xi^+_c \to p K^- \pi^+$ has been measured by FOCUS~\cite{FOCUS:2001ovr}, and the $K^-p$ invariant mass distribution show a clear peak of $\Lambda(1520)$ and a dip-like structure around $\eta\Lambda$ threshold.  The future measurements of this process in Belle and STCF could be used to determine the properties of the $\Lambda(1670)$.

Recently, Ref.~\cite{Liu:2023xvy} has studied two different scenarios for the internal structure of the $\Lambda(1670)$. One scenario assumes that the $\Lambda(1670)$ is dynamically generated through meson-baryon scattering, and the other assumes that the $\Lambda(1670)$ is a bare quark-model-like state mixing with these $I=0$ interacting channels. It is concluded that one cannot distinguish those two scenarios using scattering data alone, and exploring the structure of the $\Lambda(1670)$ in the finite volume of lattice QCD is necessary~\cite{Liu:2023xvy}. In addition, recent lattice QCD results~\cite{BaryonScatteringBaSc:2023zvt,BaryonScatteringBaSc:2023ori} for the $J^P=1/2^-$ $\Lambda$ channel implies that, including a single particle contribution in the $\Lambda(1670)$ structure is important. Thus, the internal structure of the $\Lambda(1670)$ is not well understood, and the more precise measurements could shed light on this question.

\section{ACKNOWLEDGMENTS}

We would like to thank Eulogio Oset for useful comments.
This work is supported by the National Key R\&D Program of China (No. 2024YFE0105200), the Natural Science Foundation of Henan under Grant No. 232300421140 and No. 222300420554, the National Natural Science Foundation of China under Grant No. 12205075, No. 12475086, and No. 12192263.


\end{CJK}
\end{document}